\documentclass[a4paper, amsfonts, amssymb, amsmath, preprint, showkeys, nofootinbib, twoside]{revtex4-1}
\usepackage[english]{babel}
\usepackage[utf8]{inputenc}
\usepackage[colorinlistoftodos, color=green!40, prependcaption]{todonotes}
\usepackage{amsthm}
\usepackage{mathtools}
\usepackage{physics}
\usepackage{xcolor}
\usepackage{graphicx}
\usepackage[left=23mm,right=13mm,top=35mm,columnsep=15pt]{geometry} 
\usepackage{adjustbox}
\usepackage{placeins}
\usepackage[T1]{fontenc}
\usepackage{lipsum}
\usepackage{csquotes}
\usepackage[pdftex, pdftitle={Article}, pdfauthor={Author}]{hyperref} 
\newcommand{\rev}[1]{\textcolor{black}{#1}}
\newcommand{\normord}[1]{:\mathrel{#1}:}

\begin{document}

\title{Broadband complex two-mode quadratures for quantum optics}

\author{$\text{Leon Bello}^{1,3}$, $\text{Yoad Michael}^1$, $\text{Michael Rosenbluh}^1$,  $\text{Eliahu Cohen}^2$, $\text{Avi Pe'er}^1$}
\email[Correspondence email address: ]{avi.peer@biu.ac.il}
\affiliation{1 Department of Physics and BINA Institute of Nanotechnology, Bar-Ilan University, Ramat-Gan 52900, Israel}
\affiliation{2, Faculty of Engineering, Bar-Ilan Universit, Ramat-Gan 52900, Israel}
\affiliation{3 Department of Electrical and Computer Engineering, Princeton University, Princeton, NJ}
    
\date{\today} 

\begin{abstract}
In their seminal paper, Caves and Schumaker presented a new formalism for quantum optics, intended to serve as a building block for describing two-photon processes, in terms of new, generalized qudratures. The important, revolutionary concept in their formalism was that it was fundamentally two-mode, i.e. the related observables could not be attributed to any single one of the comprising modes, but rather to a generalized complex quadrature that could only be attributed to both of them. Here, we propose a subtle, but fundamentally meaningful modification to their important work. Unlike the above proposal, we deliberately choose a frequency-agnostic definition of the two-mode quadrature, that we motivate on physical grounds. This simple modification has far-reaching implications to the formalism -- the real and imaginary parts of the quadratures now coincide with the famous EPR variables, and our two-mode operators transform trivially under two-mode and single-mode squeezing operations. Their quadratic forms, which we call the ``quadrature power'' are shown to succinctly generate the $SU(1,1)$ algebra of squeezing Hamiltonians, and correspond directly to important, broadband physical observables, that have been directly measured in experiment and are explicitly related to properties like squeezing and entanglement.
This new point of view gives a fresh perspective on two-mode processes that is completely agnostic to the bandwidth, and reveals intriguing new ways for understanding and measuring broadband two-mode squeezing.
\end{abstract}


\maketitle

\section{Introduction}
Two-photon devices, like parametric amplifiers and four-wave mixers, are the workhorse of modern quantum optics, and are of fundamental importance in all fields of quantum field theory. In the limit of small photon numbers, they are used as sources of indistinguishable photons \cite{mandel_1987}, entangled photon pairs \cite{mandel_1985, franson_1989, weinberg_simultaneity_of_photon_pairs} and heralded single photons \cite{weighs_2009, mandel_1986}. The unique and useful properties of two-mode devices stem from the creation and annihilation of quanta exclusively in the form of entangled pairs, with no quantum properties bestowed on the individual particles that comprise the pair, which appear as thermal radiation \cite{Yurke1987}. They comprise a very important resource in continuous-variable quantum optics, employed to generate squeezed-vacuum and squeezed-coherent states, which are critical for continuous-variable quantum computation and quantum information processing \cite{Braunstein2005, Gu2009}, quantum sensing and precise metrology \cite{Giovannetti2011, caves_noise_limits, michael_2019} with the notable example of gravitational wave detection \cite{Aasi2013, tse_2019}), spectroscopy, and also coherent computation schemes \cite{Marandi2014, bello_2019}. Furthermore, two-mode phenomena are at the core of fundamental processes all across physics, from theoretical black-hole thermodynamics \cite{Unruh1976, Harlow2016}, condensed-matter physics \cite{Bardeen1957, Svozil1990} and quantum information \cite{Einstein1935, Weedbrook2012} to applied experiments in circuit quantum electrodynamics and quantum optomechanics, where two-mode processes can occur between modes at vastly different domains (e.g. one optical and the other mechanical/electrical) \cite{Clerk_2010, Aspelmeyer2014, Moon2005}.

It is very common in quantum optics to describe processes using the quadrature amplitudes. These are fundamental quantities of both quantum and classical optics that appear in various applications across physics and related sciences. In quantum field theory and quantum optics, the quantized electromagnetic field modes are treated as quantum harmonic oscillators, where the quadrature amplitudes, the in-phase and in-quadrature components of the field are conjugate variables that play the roles of position and momentum, respectively, and are also important for the definition of the phase operator \cite{Susskind1964}. Even outside of physics, the quadrature amplitudes are a basic concept in communication theory and signal analysis \cite{haykin_signals_and_systems}, where they generalize the notion of amplitude and phase, and are the central building blocks in the theory of coherence \cite{goodman_statistical_optics}.  

In the case of narrowband light, the dynamics of parametric devices and amplifiers is described elegantly in terms of the quadrature amplitudes $\hat{x} = (\hat{a} + \hat{a}^\dagger)/\sqrt{2}$ and $\hat{y} = -i(\hat{a} - \hat{a}^\dagger)/\sqrt{2}$, where the operation of the degenerate parametric amplifier is simply phase-sensitive amplification of one quadrature  $\hat{x}\rightarrow \hat{x}e^r$ at the fundamental cost of attenuating the other $\hat{y}\rightarrow \hat{y}e^{-r}$, where $r$ is the (real) squeezing parameter \cite{Clerk_2010}. In the non-degenerate case, this picture becomes more complicated, and squeezing is described only in terms of the correlations between the modes.
The standard description of two-mode phenomena is therefore based on the single-mode quadrature operators of the individual modes and the correlations between them \cite{Simon2000, Duan2000, Ou1992}, rather than a unified observable that describes the pair and can be measured directly. 

\textcolor{black}{
In their famous 1985 papers \cite{caves_new_formalism_i, Schumaker1985}, Caves and Schumaker realized the importance of two-photon processes, and  suggested that since these processes involve only the creation and annihilation of photon pairs, the natural way to describe them is using a formalism that is inherently bi-modal. In their work, they defined a new pair of non-Hermitian quadrature operators that generalize and extend the quadrature description for two-mode fields by introducing a new set of non-local complex quadrature amplitude operators, for describing a pair of modes with frequency separation $\epsilon$, around a central ``carier'' mode with frequency $\Omega$,}
\begin{subequations}
    \begin{align}
          \alpha_1(\epsilon) = \left ( \frac{\Omega + \epsilon}{2\Omega} \right )^{1/2} \hat{a}(\Omega + \epsilon) + \left ( \frac{\Omega - \epsilon}{2\Omega} \right )^{1/2} \hat{a}^\dagger(\Omega + \epsilon) \\ 
          \alpha_2(\epsilon) = i\left ( \frac{\Omega + \epsilon}{2\Omega} \right )^{1/2} \hat{a}(\Omega + \epsilon) - i \left ( \frac{\Omega - \epsilon}{2\Omega} \right )^{1/2} \hat{a}^\dagger(\Omega + \epsilon)
    \end{align}
\end{subequations}
\begin{subequations}
	\begin{align}
	\left[\hat{\alpha}_{1},\hat{\alpha}_{1}^{\dagger}\right]=\left[\hat{\alpha}_{2},\hat{\alpha}_{2}^{\dagger}\right] &= \epsilon/\Omega \\
	\left[\hat{\alpha}_{1},\hat{\alpha}_{2}^{\dagger}\right]=\left[\hat{\alpha}_{1}^{\dagger},\hat{\alpha}_{2}\right] &= i \\
	\big[\hat{\alpha}_{1},\hat{\alpha}_{2}\big] &= 0
	\end{align}
\end{subequations}

\textcolor{black}{The frequency normalization of these operators was introduced to maintain their direct reflection of the observable electric field. This normalization however leads to a rather complicated algebra that depends on the energy separation between the modes. We propose a subtle change to that definition, that has far reaching consequences to the two-mode formalism. We deliberately omit the frequency dependence of the two-mode quadrature amplitudes, which dramatically changes the algebra and interpretation of the two-mode amplitudes.
Our operators are thus completely agnostic to the frequency separation between the modes. Their commutation relations are now identical to those of the canonical single-mode quadrature, and they very cleanly generate the $SU(1,1)$ algebra of parametric processes. Moreover, they directly relate to the well-known EPR observables, and succinctly demonstrate how the EPR correlations follow from the squeezing of the complex quadratures. The real and imaginary components of the complex quadratures are observable quantities, and correspond directly to the famous EPR correlations.}
\textcolor{black}{The above has interesting implications for two-mode parametric processes and for ultra broadband applications thereof. The altered quadratures describe bosonic pairs regardless of their frequency separation, allowing us to define global observables that can be attributed to signals of any bandwidth.} 

\textcolor{black}{Another helpful concept that is introduced in our work is that of the quadrature powers -– since we represent the quadratures using complex phasors, it is natural to ascribe a power operator to them.}
We define a set of Hermitian operators based on the complex quadratures - the quadrature power operators. These generate the $SU(1,1)$ algebra of two-photon devices and correspond closely to the observables of the $SU(1,1)$ interferometer \cite{Chekhova2016, Yurke1986}, which measures the nonlinear interference between two conjugate signal-idler fields of a two-mode squeezed state. Moreover, using the complex-quadrature formalism, we expand  previous experimental results \cite{Shaked2018}, and offer a novel measurement scheme closely related to $SU(1,1)$ interferometers. This formalism offers two key advantages: First, it offers a method to measure inseparability and squeezing using average power measurement instead of coherent homodyne detection. Second, it enables the use of non-ideal detectors, and possesses an effectively unlimited bandwidth.
Additionally, it also presents a solution to a fundamental asymmetry in parametric processes - it is quite easy to generate very broadband fields using processes like spontaneous down-conversion, but it is extremely difficult to make broadband measurements of them by electronic means. The available very large bandwidth is an untapped resource that can be easily generated, but cannot be used. As we show in this work, the quadrature power operators are slowly varying observables that can be efficiently measured over any bandwidth, solving the above asymmetry, and paving the way towards applications that exploit the broadband capabilities of parametric processes. 

\textcolor{black}{While the change in definition is simple, a straightforward omission of the frequency normalization in Caves and Schumaker's original formulation, the implications to the algebra, features, concepts, and applications of the formalism are far-reaching, and completely absent from Caves and Schumaker’s original formalism.}

\textcolor{black}{Caves and Schumaker briefly consider a non-frequency normalized definition of the complex quadratures, but quickly dismiss it as unphysical, and do not explore them beyond the basic definition.}
\textcolor{black}{They claim, following a paper by Kimble and Mandel \cite{mandel_polychromatic_detection}, that these operators do not correspond to what is physically measured in homodyne detection, and thus are not physically meaningful. Kimble et al. \cite{mandel_polychromatic_detection} discuss the problem of photoelectric detection of polychromatic, non-stationary light and show that  it cannot be argued that the photodetector measures the photon flux. They also show that in the limit of monochromatic light, the direct relation between photon flux and the detector photo-current is retrieved. This critique applies to direct detection, but not to the two-mode schemes and devices discussed in this work, e.g. two-photon processes like sum- and difference-frequency generation.}
Another argument raised in the original works of Caves et al., which essentially relates to the same point, is that the single mode operators do not have the same units; each operator has units of the square-root of the number of photons with respect to its own frequency, and thus they cannot be summed together without proper normalization. 
We respectfully disagree with their conclusion, and claim that in fact, this definition correctly captures the relevant physics and observables of two-mode devices. We believe this claim is not applicable for two-mode processes, since their defining property, as shown by Manley and Rowe \cite{Manley_1956, rowe_ii} is that they depend on the \emph{number of quanta in each mode}, with respect to its own frequency, and not with respect to a common carrier.

The paper is organized as follows: In \textbf{section \ref{sec:complex_quadratures}} we introduce the two-mode quadratures formalism. We first give a classical motivation for our definition, and then introduce the operator-valued counterparts. We explore their commutation relations and algebraic structure, and relate it to the well known $SU(1,1)$ algebra. In \textbf{section \ref{sec:two-mode_devices}} we apply our formalism to two-photon devices, and show how the two-mode quadratures make the natural variables for describing them and how their unique properties are elucidated. In \textbf{section \ref{sec:experimental_significance}} we discuss the applicative significance of our formalism and suggest experimental schemes that illuminate and exploit the complex quadratures. Finally, in \textbf{section \ref{sec:conclusions}} we conclude and revisit the theoretical and practical merits of the picture presented in this paper.
\section{Results}
	
	\subsection{The complex quadrature operators}
	\label{sec:complex_quadratures}
	
	\subsubsection{Complex two-mode quadrature operators}
	The quadrature amplitudes are the principal observables in the continuous-variable description of quantum fields.
	\begin{subequations}
		\begin{align}
			\hat{x}_\omega = \frac{1}{\sqrt{2}} \left( \hat{a}_\omega  + \hat{a}_\omega^\dagger \right) \\
			\hat{y}_\omega = \frac{1}{\sqrt{2} i} \left( \hat{a}_\omega - \hat{a}_\omega^\dagger \right) \\
			\left[ \hat{x}_\omega, \hat{y}_{\omega'} \right] = i \delta_{\omega,\omega'}.
		\end{align}
	\end{subequations}
	where the quadratures are normalized in the common quantum-optical convention such that their commutator is equal to $i$. 
	
	Their importance stem from the ability to directly measure them by homodyne detection, but measurements of the quadrature amplitudes are usually thought to be meaningful only in the monochromatic limit, close to the degenerate limit where the complex envelope varies on slow timescales easily distinguishable from the carrier, and detectable by electronic photo-detectors. Our goal here is to generalize the quadrature operators for two-mode signals of any bandwidth and form, and show that even extremely broadband signals could still be measured in a simple and physically meaningful way. 
	
	In what follows, we consider parametric two-mode devices, which have the specific property that they create and annihilate photons exclusively in pairs of conjugate modes (referred to as the signal and idler). \textcolor{black}{For concreteness, we focus on spectral modes}, and consider conjugate frequency modes, but the formalism can be readily applied to any multimode description, e.g. two-mode spatial fields with conjugate momenta.
	
	Let us consider a pair of frequency modes $\omega_{\pm} = \Omega \pm \epsilon$, symmetrically separated from a central ``carrier'' mode at frequency $\Omega$ by $\epsilon$. These modes are commonly referred to as ``signal'' and ``idler'' modes, respectively. The full Hamiltonian of the field including the parametric interaction, assuming a classical pump field with negligible depletion \textcolor{black}{in the carrier frame}, is given by
	\begin{equation}
		\label{eq:two-mode_freq}
		\hat{H} = \hbar \epsilon \left( \hat{a}_+^\dagger \hat{a}_+ - \hat{a}_-^\dagger \hat{a}_- \right)
		+ i\left(\xi^{*}a_{+}a_{-}-\xi a_{+}^{\dagger}a_{-}^{\dagger}\right)
	\end{equation}
	where $\hat{a}_{\pm}=\frac{1}{\sqrt{2}}\left(\hat{x}_{\pm}+i\hat{y}_{\pm}\right)$ are the mode operators, satisfying the canonical commutation relations $\left[\hat{a}_{i},\hat{a}_{j}^{\dagger}\right]=\delta_{ij}$. In such two-photon processes, pump photons are directly converted into pairs of signal and idler quanta. \textcolor{black}{Since the process adds the same number of photons to each mode, the photon-number difference is conserved}:
	\begin{equation}
		\label{eq:number_difference}
		\hat{\delta N} = \hat{a}_+^\dagger \hat{a}_+ - \hat{a}_-^\dagger \hat{a}_-.
	\end{equation}
	This is the defining symmetry of parametric processes, \textcolor{black}{and the main motivator to our frequency-agnostic definition} -- what matters is not the power, but the number of photons in each mode. \textcolor{black}{Parametric processes conserve the photon-number difference between the signal and idler modes}, in contrast to the standard beam splitter, which conserves total photon-number $\hat{N} = \hat{a}_+^\dagger \hat{a}_+ + \hat{a}_-^\dagger \hat{a}_-$. Parametric interactions therefore follow the $SU(1,1)$ symmetry group (as opposed to linear optics, which follows $SU(2)$), hence the name $SU(1,1)$ interference \cite{Yurke1986}. Classically, this follows from the well-known Manley-Rowe relations \cite{rowe_ii} applied to parametric processes. Parametric devices amplify the field as conjugate pairs of modes, as captured by the generalized Bogoliubov transformation
	\begin{equation}
		\label{eq:bogoliubov}
		\hat{a}_\pm \rightarrow \hat{a}_\pm \cosh r +  \hat{a}_\mp^\dagger \sinh r 
	\end{equation}
	where $r = \int_{0}^{t} \frac{\xi}{\hbar} \textrm{dt}$ \textcolor{black}{is the squeezing strength}. Classically, this follows from the phase-conjugation of parametric processes like difference-frequency generation and four-wave mixing. The conservation of the difference in the number of photons in Eq. \ref{eq:number_difference}. The conjugate pair amplification in Eq. \ref{eq:bogoliubov} stems from the inherent $SU(1,1)$ symmetry of two-mode processes, and is true for any bandwidth and any carrier frequency.
	
	In the degenerate case, \textcolor{black}{the quadrature operators transform very simply under squeezing -- one operator is amplified, $\hat{x}e^r$, while the other is attenuated, $\hat{y}e^{-r}$. Note that it is the operators themselves that transform, and not just the expectation values, which implies that all moments transform trivially and noiselessly under the squeezing transformation.}
	\textcolor{black}{Extending that perspective, we can think of parametric amplification as a process that amplifies the real part of the field, and attenuates the imaginary part, where the orientation of the real and imaginary axes is defined by the pump frequency and phase.}
	
\begin{figure}
		\centering
		\includegraphics[width=8.6cm]{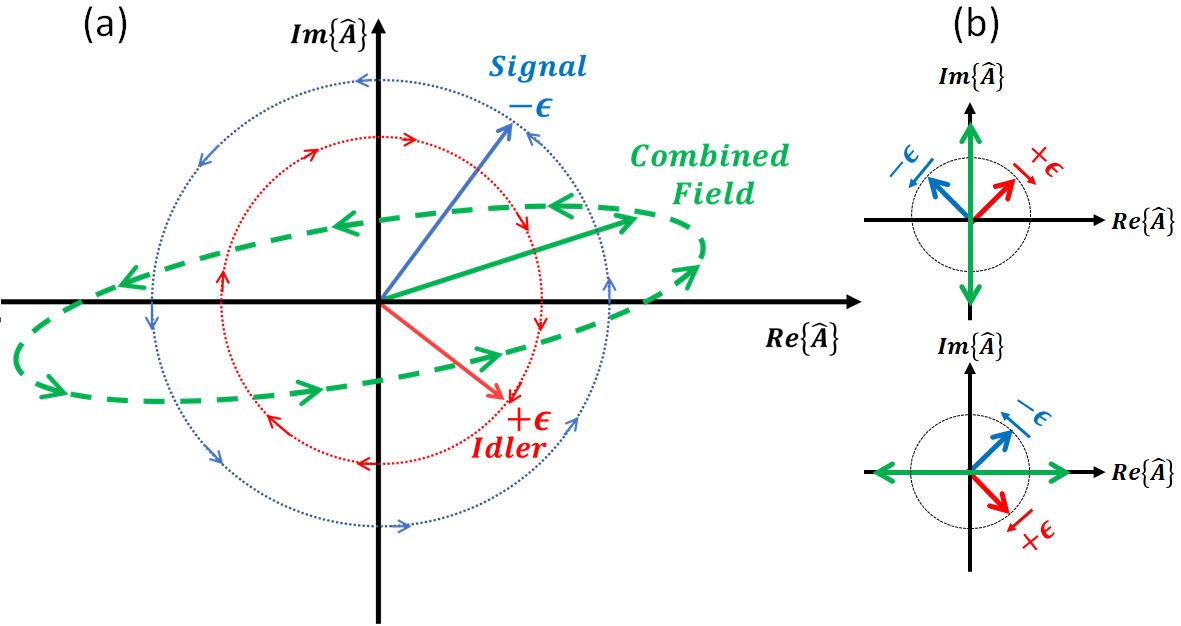}
		\caption{\textbf{(a)} The quadrature map of a general two-mode field, as observed in a rotating frame at the center frequency. The signal (idler) field is denoted by the blue (red) arrow. In this frame the signal (idler) rotates counter-clockwise (clockwise) at the offset frequency from the carrier $\epsilon$. The total two-mode field (green arrow) is therefore harmonically oscillating along the marked elliptical trajectory. \textbf{(b)} shows the special case of equal amplitudes and conjugate phases for the signal and idler, where the two-mode elliptical oscillation trajectory collapses to a single line along one of the axes - the definition of a pure two-mode quadrature.}
		\label{fig:time_dependent_quadratures}
	\end{figure}

	Let us employ a similar perspective for the non-degenerate case, and try to similarly divide the field it into a real and imaginary part. We look at the complex amplitudes in the rotating frame of the carrier $\Omega$, as illustrated in Fig. \ref{fig:time_dependent_quadratures}, the two modes rotate in opposite directions with opposing frequencies $\pm\epsilon$, causing the total field amplitude (and its corresponding quadrature amplitudes) to \textcolor{black}{oscillate} in time \textcolor{black}{(beat)} at the difference frequency between the modes.
	
	\textcolor{black}{As we have established before, the parametric amplifier performs two actions -- it adds the same number of photons into both modes (conserving the photon-number difference), and conjugates them. If we think of a generalized two-mode creation operator, which incorporates both fields, these two properties together amount to the amplification of the modes in such a way that the real part of the total field is amplified, while the imaginary part is attenuated, exactly in accordance with the degenerate case, as illustrated in \ref{fig:time_dependent_quadratures}.}
	
	Let us now define the quantum field operator for the total, time-dependent complex amplitude of the two-mode field as
	\begin{equation}
		\hat{A}(t) = \left( \hat{a}_{+}e^{-i\epsilon t} + \hat{a}_{-}e^{+i\epsilon t} \right) e^{-i\Omega t}.
	\end{equation}
	\textcolor{black}{Here starts the major deviation from Caves and Schumaker's definition -- we do not look at the real part of the field, but of a generalized two-mode annihilation operator. As discussed,} according to Eqs. \ref{eq:two-mode_freq}, \ref{eq:number_difference} and \ref{eq:bogoliubov}, the parametric amplifier simply amplifies the real part of the total time-dependent complex envelope, and attenuates the imaginary part, \textcolor{black}{regardless of the time-dependence and bandwidth of the two-mode complex envelope. This allows us to generalize the single-mode quadratures, and define a pair of complex two-mode quadrature amplitudes $\hat{X}_1$ and $\hat{X}_2$ as the real and imaginary parts of the two-mode complex envelope around the carrier frame defined by $\Omega$. As we shall see, this definition is convenient since it behaves exactly the same for both degenerate and non-degenerate cases, and many of the properties of squeezed-states stem directly from it.}
	\begin{equation}
		\hat{X}_1 (t) \equiv \frac{1}{\sqrt{2}}\left[ \left(x_{+}+x_{-}\right)\cos\left(\epsilon t\right)+\left(y_{+}-y_{-}\right)\sin\left(\epsilon t\right) \right]
	\end{equation}
	\begin{equation}
		\hat{X}_2 (t) \equiv \frac{1}{\sqrt{2}} \left[ \left(y_{+}+y_{-}\right)\cos\left(\epsilon t\right)-\left(x_{+}-x_{-}\right)\sin\left(\epsilon t\right) \right],
	\end{equation}
	where the operators $\hat{x}_{\pm}$ and $\hat{y}_{\pm}$ are the standard single-mode quadrature amplitudes of each mode. Removing the trivial time-dependence on the beat frequency $\epsilon$ and the carrier $\Omega$, we obtain the complex two-mode quadrature operators,
	\begin{subequations}
		\label{eq:complex_quadratures_1}
		\begin{align}
			\hat{X}_{1}=\frac{1}{\sqrt{2}}\left(\hat{a}_{+}+\hat{a}_{-}^{\dagger}\right) =\frac{1}{2}\left[\left(\hat{x}_{+}+\hat{x}_{-}\right)+i\left(\hat{y}_{+}-\hat{y}_{-}\right)\right] \\
			\hat{X}_{2} = \frac{1}{\sqrt{2}i}\left(\hat{a}_{+}-\hat{a}_{-}^{\dagger}\right) =   \frac{1}{2}\left[\left(\hat{y}_{+}+\hat{y}_{-}\right)-i\left(\hat{x}_{+}-\hat{x}_{-}\right)\right].
		\end{align}
	\end{subequations}
    To avoid confusion, we denote throughout this manuscript the two-mode complex quadrature amplitudes with capital letters $\hat{X}_{1(2)}$ and numeric subscripts, and the single-mode quadrature amplitudes $\hat{x}_\pm$ and $\hat{y}_\pm$ with lower-case letters and non-numeric subscripts. Also note that the $\pm$ subscript refers to the individual modes, and not to sums and differences of the positions and momenta as is sometimes used in the literature of two-mode squeezed states.
    \textcolor{black}{It is also interesting to note that in the narrowband limit $\epsilon\ll\Omega$, we approach the single-mode description of a degenerate parametric amplifier and the quadrature amplitudes can be thought of as the slowly varying amplitudes of the cosine and sine parts of the carrier, further motivating the definition.}
    
    \begin{figure}
		\centering
		\includegraphics[width=8.6cm]{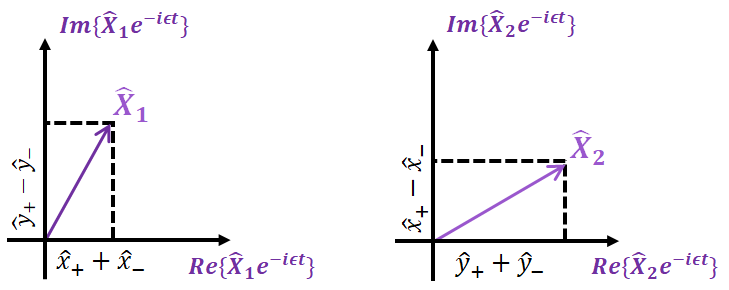}
		\caption{The complex quadratures:  Since the two-mode quadratures are harmonically oscillating in time at the beat frequency $\epsilon$, they are conveniently represented as complex amplitudes (in a rotating frame at frequency $\epsilon$, where the real and imaginary components correspond to the well known two-mode correlations between the single-mode homodyne measurements shown in Eq. \ref{eq:quadrature_components_1}-\ref{eq:quadrature_components_2} and \ref{eq:quad_re_im}.}
		\label{fig:complex_quadratures}
	\end{figure}
	
    \textcolor{black}{Note that the real and imaginary parts of the complex quadrature amplitudes}
    \begin{subequations} \label{eq:quad_re_im}
		\begin{align}
			\hat{X}_{1}=\frac{1}{\sqrt{2}}\left(\hat{\chi}_{1}+i\hat{\gamma}_{1}\right) \\
			\hat{X}_{2}=\frac{1}{\sqrt{2}}\left(\hat{\chi}_{2}+i\hat{\gamma}_{2}\right)
		\end{align}
	\end{subequations}
 \textcolor{black}{can be directly identified as the EPR variables}	
	\begin{subequations}
		\label{eq:quadrature_components_1}
		\begin{align}
			\hat{\chi}_{1}&\equiv\frac{1}{\sqrt{2}}\left(\hat{x}_{+}+\hat{x}_{-}\right) \\
			\hat{\gamma}_{1}&\equiv\frac{1}{\sqrt{2}}\left(\hat{y}_{+}-\hat{y}_{-}\right)
		\end{align}
	\end{subequations}
	This gives a clear experimental meaning to the complex quadratures - their components are the EPR observables, and can be measured directly in a two-mode homodyne experiment \cite{Li2019}. 
	
	\begin{subequations}
		\label{eq:quadrature_components_2}
		\begin{align}
			\hat{\chi}_{2}&\equiv\frac{1}{\sqrt{2}}\left(\hat{y}_{+}+\hat{y}_{-}\right) \\
			\hat{\gamma}_{2}&\equiv-\frac{1}{\sqrt{2}}\left(\hat{x}_{+}-\hat{x}_{-}\right).
		\end{align}
	\end{subequations}
	From the commutation relations we derived above, one can show that
	\begin{subequations}
		\begin{align}
			\left[\hat{\chi}_1,\hat{\chi}_2\right] &= \left[\hat{\gamma}_1, \hat{\gamma}_2\right]=i \\
			\left[\hat{\chi}_i, \hat{\gamma}_j\right] &= i\varepsilon_{ij}	
		\end{align}
	\end{subequations}
	where $\varepsilon_{ij}$ is the two-dimensional Levi-Civita symbol. \textcolor{black}{Importantly, referring to Eqs. \ref{eq:quadrature_components_1} and \ref{eq:quadrature_components_2}, we notice that indeed, although the complex quadratures are non-Hermitian, they are normal operators and their components commute and make two simultaneously measurable observables.}
	\textcolor{black}{This directly relates squeezing to the EPR observables, a notion that does not follow from Caves and Schumaker's original frequency-dependent definition, but follows very naturally from ours.}
	
	Similarly to the single-mode quadrature operators, it is useful to define a general quadrature operator, generated by the Hamiltonian $\hat{H}_\Omega = \hbar\Omega\left( \hat{a}_+^\dagger\hat{a}_+ + \hat{a}_-^\dagger\hat{a}_- \right)$, at an arbitrary quadrature axis rotated by an angle $\theta$ relative to the previous quadrature axes. The two-mode quadratures are defined with respect to the ``carrier-phase'' associated with the modes $\theta = \frac{1}{2}(\theta_{+} + \theta_{-}) $, where $\theta_{\pm}$ denotes that phase of the mode $\hat{a}_{\pm}$
	\begin{equation}
		\begin{array}{c}
			\hat{X}_{1}\left(\theta\right) \equiv \hat{X}_{1}\cos\theta+\hat{X}_{2}\sin\theta \\
			\hat{X}_{2}\left(\theta\right)\equiv \hat{X}_{1}\left(\theta + \frac{\pi}{2}\right) = - \hat{X}_{1}\sin\theta + \hat{X}_{2}\cos\theta ,
		\end{array}
	\end{equation}  
	which is very similar to the single-mode definition, and indeed coincides with it in the degenerate limit $\omega_+ = \omega_-$. As is emphasized throughout this work - the complex quadratures generalize the single-mode quadratures, and treat on equal footing both degenerate and non-degenerate two-photon processes.
	
	The complex quadrature operators satisfy a set of very elegant commutation relations. First, $\hat{X}_{1}$ and $\hat{X}_{2}^\dagger$ comprise a conjugate pair, analogous to the single-mode quadrature operators
	\begin{equation}
		\left[\hat{X}_{1}, \hat{X}_{2}^{\dagger}\right]=\left[\hat{X}_{1}^{\dagger}, \hat{X}_{2}\right] = i.
	\end{equation}
	Although with an important distinction - they operators are not Hermitian, and thus $\hat{X}_i$ and $\hat{X}^\dagger_i$ are different operators. In the degenerate case, the quadrature operators become Hermitian again, and we retrieve the standard canonical relation.
	$\hat{X}_i$ is a diagonalizable normal operator, and$\hat{X}_{1}$ and $\hat{X}_{2}$ commute and each of them commutes with its conjugate
	\begin{equation}
		\left[\hat{X}_{1},\hat{X}_{2}\right] = 0
	\end{equation}
	\begin{equation}
		\label{eq:complex_commutation_relations}
		\left[\hat{X}_{1},\hat{X}_{1}^{\dagger}\right]=\left[\hat{X}_{2},\hat{X}_{2}^{\dagger}\right] = 0
	\end{equation}
	The complex quadratures satisfy the same commutation relations as the canonical quadratures, \textcolor{black}{another major result that stems from our definition}. Similarly, they have a set of unnormalizable eigenfunctions - the famous EPR states as we show in Sec. \ref{sec:two-mode_devices}.
	
	Writing the Heisenberg equations of motion for the interaction Hamiltonian, we find that under the operation of a squeezer $\hat{S}(r) = e^{-\left(r^{*}a_{+}a_{-}-ra_{+}^{\dagger}a_{-}^{\dagger}\right)}$ along the same principal axes, the complex quadrature operators change very simply -- \textcolor{black}{they are multiplied by a scale factor:}
	\begin{subequations}
		\begin{align}
			\hat{S}^\dagger_1(r) \hat{X}_1 \hat{S}_1(r) = e^{+r}\hat{X}_1 \\
			\hat{S}^\dagger_1(r) \hat{X}_2 \hat{S}_1(r) = e^{-r}\hat{X}_2
		\end{align}
	\end{subequations}
	While the two-mode (non-degenerate) parametric amplification necessarily adds noise to the single-mode quadratures, the complex two-mode quadratures incur no such noise, completely analogous to the degenerate case.
	
	\subsubsection{Quadrature power and quadrature coherence operators}
	The real and imaginary parts of the complex quadratures commute and can be measured simultaneously in a coherent two-mode homodyne scheme \cite{Porta1989, Li2019}.
	However, we are often interested only in the second moments of the quadrature amplitudes \cite{Ou1992, Simon2000, Duan2000}, for example when measuring Gaussian states.
	In experimental schemes, it is often convenient to make average power measurements, rather than coherent detection. In this section, we introduce the power and correlation operators of the two-mode quadratures. These allow us to generalize the two-mode correlations, and reveal a convenient measurement using only parametric amplifiers and direct detection. These quantities are directly measurable experimentally and play a central role in the interpretation of two-photon states, such as squeezed-states. In Sec. \ref{sec:experimental_significance}, we show that the quadrature powers are measured by photo-detection at the output of a parametric amplifier.
	
	Let us first define the following symmetric matrix of Hermitian operators,
	\begin{equation}
		\hat{\Gamma}_{ij} = \frac{1}{2} \left( \hat{X}_i^\dagger \hat{X}_j + \hat{X}_j^\dagger \hat{X}_i \right),
	\end{equation}
	where $\hat{\Gamma}_{ij}$ is the quadrature operator matrix. The diagonal elements of this matrix represent the power associated with each quadrature, whereas the off-diagonal elements represent the correlations between the quadratures. 
	Let us elaborate on the diagonal operators - the quadrature powers, $\hat{\Gamma}_{11}$ and $\hat{\Gamma}_{22}$:
	\begin{subequations}
		\begin{align}
			\hat{\Gamma}_{11} = \hat{X}_1^\dagger \hat{X}_1 = \frac{1}{2} \left(\hat{a}_{+}^\dagger \hat{a}_{+} + \hat{a}_{+}^\dagger \hat{a}_{-}^\dagger + \hat{a}_{+} \hat{a}_{-} + \hat{a}_{-} \hat{a}_{-}^{\dagger} \right) \\
			\hat{\Gamma}_{22} = \hat{X}_2^\dagger \hat{X}_2 = \frac{1}{2} \left(\hat{a}_{+}^\dagger \hat{a}_{+} - \hat{a}_{+} \hat{a}_{-} - \hat{a}_{+}^\dagger \hat{a}_{-}^\dagger + \hat{a}_{-} \hat{a}_{-}^{\dagger} \right),
		\end{align}
	\end{subequations}
	The phase-dependent properties of parametric devices are encoded in the correlations $\hat{a}_+ \hat{a}_-$, which are important observables for various applications, e.g. quantum illumination \cite{Zhuang2017, Zhang2015}. Tracing over $\hat{\Gamma}_{ij}$ gives the total number of photons up to normal ordering
	\begin{equation}
		\label{eq:total_power}
		\hat{N} = \ \normord{\text{tr}(\hat{\Gamma})} \  = \normord{\hat{\Gamma}_{11}} + \normord{\hat{\Gamma}_{22}} = \hat{a}_{+}^\dagger \hat{a}_{+} + \hat{a}_{-}^\dagger \hat{a}_{-} .
	\end{equation}
	
	The non-diagonal element, i.e. the quadrature coherence, represents the coherence or correlations between the two quadrature amplitudes:
	\begin{equation}
		\hat{\Gamma}_{12} = \frac{1}{2} \left( \hat{X}_1^\dagger \hat{X}_2 + \hat{X}_2^\dagger \hat{X}_1 \right) = \frac{1}{2} \left[ \hat{\Gamma}_{22}^{(\theta - \pi/4)} - \hat{\Gamma}_{11}^{(\theta - \pi/4)}\right] \\
		= \frac{i}{2} \left( \hat{a}_+^\dagger \hat{a}_-^\dagger - \hat{a}_- \hat{a}_+ \right).
	\end{equation}
	Expressed in the orthogonal basis, where the quadrature operators are rotated by $\pi/4$, the quadrature coherence corresponds simply to the difference in the quadrature powers.
	Together with the quadrature powers, they determine both the real and imaginary parts of the intermodal coherence, which encodes the quantum signature of squeezed-light \cite{Zhuang2017, Zhang2015, Sanz2017}.
	
	The quadrature powers satisfy a closed set of commutation relations:
	\begin{subequations}
		\begin{align}
			\left[\hat{\Gamma}_{11}, \hat{\Gamma}_{22}\right] &= 2i \hat{\Gamma}_{12} \\
			\left[\hat{\Gamma}_{22}, \hat{\Gamma}_{12}\right] &= -i \hat{\Gamma}_{22} \\
			\left[\hat{\Gamma}_{12}, \hat{\Gamma}_{11}\right] &= -i \hat{\Gamma}_{11}.
		\label{eq:quadrature_powers_squeezing}
		\end{align}
	\end{subequations}
	In general, they are not commuting observables. For stationary states with uniform phase noise, such as coherent states, $\left\langle\hat{\Gamma}_{12}\right\rangle=0$ and the quadrature powers can be measured simultaneously. For non-stationary states, whose noise depends on the measured phase, such as squeezed states, $\left\langle\hat{\Gamma}_{12}\right\rangle\neq0$, the non-commutativity depends on the squeezing parameter. 
	
	\textcolor{black}{Since the quadrature operators just scale under squeezing, that simple scaling carries over to the quadrature powers}
	\begin{subequations}
		\begin{align}
			\hat{S}^\dagger_1(r) \hat{\Gamma}_{11} \hat{S}_1(r) &= e^{+2r}\hat{\Gamma}_{11} \\
			\hat{S}^\dagger_1(r) \hat{\Gamma}_{22} \hat{S}_1(r) &= e^{-2r}\hat{\Gamma}_{22} \\
			\hat{S}^\dagger_1(r) \hat{\Gamma}_{12} \hat{S}_1(r) &= \hat{\Gamma}_{12},
		\end{align}
	\end{subequations}
	thereby indicating that a parametric amplifier attenuates (squeezes) one quadrature power while amplifying (stretching) the other.
	This is an important result, as the quadrature powers $\hat{\Gamma}_{11}$ and $\hat{\Gamma}_{22}$ can be measured directly as we discuss later in Sec. \ref{sec:experimental_significance}. In addition, we see that the quadrature coherence $\hat{\Gamma}_{12}$ is unaffected by squeezing along the principal axes.
	
	An important application of the formalism is that, \textcolor{black}{since it is frequency agnostic}, it gives a very succinct description of broadband multimode fields, which can be easily extended to a continuum description, as given in Appendix \ref{sec:appendix}.
	
	\subsection{Description of two-mode devices and squeezed states}
	\label{sec:two-mode_devices}
	\subsubsection{Two-Mode Hamiltonian and parametric amplifiers}
	With the basic formalism defined, we now turn to employ the formalism to describe two-mode devices, i.e. processes that create and annihilate pairs of entangled photons. These two-mode interactions are described by an interaction Hamiltonian of the form 
	\begin{equation}
		\hat{H}=i\left(\xi\hat{a}_{+}\hat{a}_{-}-\xi^{*}\hat{a}_{+}^{\dagger}\hat{a}_{-}^{\dagger}\right)
	\end{equation}
	where $r=|r|e^{-i\theta}$ is the complex squeezing parameter. Experimentally, $r$ is determined by the pump field, where the phase of the pump determines which quadrature is amplified, and may in general be function of time.
	
	Two-mode interactions (such as parametric amplification) are more succinctly expressed in terms of the quadrature power operators, and can be used to write the two-mode Hamiltonian:
	\begin{equation}
		\hat{H}=-2|r|\left[\frac{1}{2}\left(\hat{\Gamma}_{11}-\hat{\Gamma}_{22}\right)\sin\theta-\hat{\Gamma}_{12}\cos\theta\right]
	\end{equation} 
	The fundamental properties of two-mode devices are captured by the $SU(1,1)$ operator algebra, \textcolor{black}{as shown in the seminal work of Yurke \cite{Yurke1986}. The quadrature powers lend themselves to very succinctly capture the $SU(1,1)$ algebra, in a way that} further emphasizes their physical significance. Let us define the following three Hermitian operators,
	\begin{subequations}
		\begin{align}
			\hat{K}_{1} &= \frac{1}{2}\left(\hat{\Gamma}_{11}-\hat{\Gamma}_{22}\right) \\
			\hat{K}_{2} &= \hat{\Gamma}_{12} \\
			\hat{K}_{3} &= \frac{1}{2}\left(\hat{\Gamma}_{11}+\hat{\Gamma}_{22}\right)
		\end{align}
	\end{subequations}
	where $\hat{\sigma}_3 = 
	\begin{pmatrix}
		1 & 0 \\
		0 & -1
	\end{pmatrix}$ is the third Pauli matrix.
	\begin{subequations}
		\begin{align}
			\left[\hat{K}_{1},\hat{K}_{2}\right] &= \frac{1}{2}\left(\left[\hat{\Gamma}_{11},\hat{\Gamma}_{12}\right]+\left[\hat{\Gamma}_{12},\hat{\Gamma}_{22}\right]\right)= i\hat{K}_{3} \\
			\left[\hat{K}_{2}, \hat{K}_{3}\right] &= \frac{1}{2}\left[\hat{\Gamma}_{12},\hat{\Gamma}_{11}+\hat{\Gamma}_{22}\right]=-i\hat{K}_{1} \\
			\left[\hat{K}_{3},\hat{K}_{1}\right] &= -\frac{1}{2}\left[\hat{\Gamma}_{11},\hat{\Gamma}_{22}\right]=-i\hat{K}_{2}
		\end{align}
	\end{subequations}
	The generators $\hat{K}_i$ of squeezing and rotation of the two-mode field are the sum and difference of the quadrature powers. $\hat{K}_{1} \propto H_{\theta = 0}$ and $\hat{K}_{2} \propto H_{\theta = \pi/2}$ are the two-mode parametric interaction terms and generate squeezing along two independent axes, \textcolor{black}{oriented at $\pi/4$  relative to each other}, while $\hat{K}_3$ is the total photon-number and generates a rotation of the quadratures due to the carrier phase. 
	This is a satisfying result, \textcolor{black}{since these generators directly correspond to real physical observables} -- $\hat{K}_{1(2)}$ are proportional to the visibility of the interference fringes in an $SU(1,1)$ interferometer along two independent squeezing axes. 
	
	\subsubsection{Conserved quantities}
	\textcolor{black}{It is well known that} the symmetry of the two-mode devices incurs a constant of motion, this is captured by the Casimir operator
	\begin{equation}
		\label{eq:casimir_invariant}
		\hat{K}^2 = \hat{K}_3^2 - \hat{K}_2^2 - \hat{K}_1^2 = \frac{1}{2}\left(\hat{\Gamma}_{11}\hat{\Gamma}_{22}+\hat{\Gamma}_{22}\hat{\Gamma}_{11}\right)-\hat{\Gamma}_{12}^{2}=\det\hat{\Gamma}
	\end{equation}
	the difference between the product of quadrature powers and the quadrature coherence is thus a conserved quantity, while the quadrature powers themselves are not. Note that since the matrix elements are operators, the determinant is an operator as well. 
	
	\textcolor{black}{This can be given an interesting physical interpretation, where $\hat{K}^2$ plays the role of the uncertainty ellipse} of the state in $\hat{\Gamma}$ space is conserved -- if one axis is squeezed, the other must be stretched. This implies the squeezing of the quadrature powers and $\hat{K}^2$ will be positive or negative depending on which axis the state is squeezed along.
	
	We defined above the symmetric operator matrix $\hat{\Gamma}$, and showed how its components generate the entire $SU(1,1)$ algebra of two-mode states and capture the coherent correlations between the comprising single-modes. However, that description is still lacking - this alone does not yet complete the description of two-mode states. The cross-quadrature coherence $X_1^\dagger X_2$ is complex, and so far we have neglected its imaginary part. As we show now, this part of the cross-quadrature coherence is precisely the difference in the number of photons between the modes. We define the anti-symmetric operator matrix
	\begin{equation}
		\hat{\Delta}_{ij} = \frac{1}{2 i} \left( \hat{X}_i^\dagger \hat{X}_j - \hat{X}_j^\dagger \hat{X}_i \right)
	\end{equation}
	The diagonal terms vanish $\hat{\Delta}_{11} = \hat{\Delta}_{22} = 0$, while the off-diagonal terms are anti-symmetric $\hat{\Delta}_{21} = -\hat{\Delta}_{12}$. In terms of the single-mode operators,
	\begin{equation}
		\hat{\Delta}_{21} = \frac{1}{2} \left(\hat{a}_+^\dagger \hat{a}_+ - \hat{a}_- \hat{a}_-^\dagger\right)
	\end{equation}
	which is the difference in the photon number in each mode, up to normal ordering. Inspection of its commutators with $\hat{\Gamma}$ shows that it commutes with all of them, and is thus a conserved quantity of a parametric process
	\begin{equation}
		\left[\hat{\Delta}_{21}, \hat{\Gamma}_{11} \right] = \left[\hat{\Delta}_{21}, \hat{\Gamma}_{22} \right] = \left[\hat{\Delta}_{21}, \hat{\Gamma}_{12} \right] = 0
	\end{equation}
	
	\textcolor{black}{To summarize, our frequency-agnostic definition of the complex quadratures allows us to define physically meaningful Hermitian operators quadratic in the complex quadratures. These operators have a plausible physical interpretation, relatively simple commutators, and they very clearly generate the squeezing algebra.}
	
	\subsubsection{Measures of inseparability and non-classical correlations}
	With the formalism laid out, we can use it to derive previous results and ideas and expand on them \textcolor{black}{for devising new applications. Especially, the quadrature powers give a very convenient way to understand and measure squeezing.}
	\begin{figure}
		\centering
		\includegraphics[width=8.6cm]{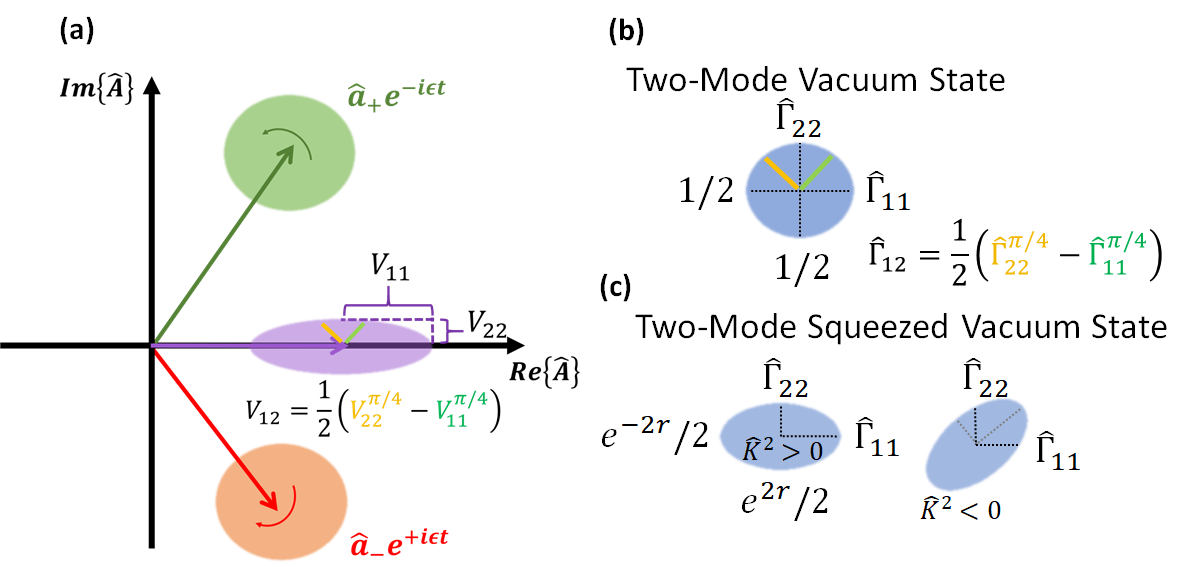}
		\caption{The ``uncertainty'' ellipse of a two-mode squeezed state. \textbf{(a)} We can draw a ``distribution'' of the complex quadrature amplitudes, representing the shared two-mode fluctuations in the state. The blue ellipse represents the distribution of the complex envelopes, and the quadrature coherence $\hat{\Gamma}_{12}$ is determined by the difference in quadrature powers along axes rotated by $\pi/4$. \textbf{(b)} If the modes are uncorrelated, like in a two-mode coherent state (or vacuum), the distribution is symmetric and does not depend on which quadrature we measure. \textbf{(c)} For squeezed states, the fluctuations depend on which complex quadrature we measure, resulting in an elliptical noise distribution.}
		\label{fig:quadrature_power_diagram}
	\end{figure}
	In the complex quadratures formalism, the non-classical entanglement of two-mode squeezed states can be described in terms of the reduction of one quadrature power below that of a vacuum state, completely analogous to degenerate squeezing.
	
	Two-mode states are inseparable -- information about the state can only be achieved by measuring both modes, while independent measurements on each mode on its own gives only partial information, or none at all \cite{Yurke1987, Barnett1988}. The state of each mode on its own is noisy and mixed, but the entangled modes are correlated better than the shot-noise limit \cite{Reid1988}. \textcolor{black}{In the complex qudarature picture, this is understood as the noiseless amplification of the complex two-mode quadrature, which at the same time leads to unitary scrambling of the single-mode quadrature.} 
	
	In single-mode squeezed states, like those generated by degenerate parametric oscillators below threshold, the state is composed only of pairs of photons whose fluctuations interfere destructively in one quadrature, and constructively in the other. This leads to increased noise along one quadrature amplitude, and decreased noise along the orthogonal one. The non-classicality of the state is manifested in the reduction of the noise fluctuations below the shot-noise limit, and the sub-Poissonian counting statistics of the state \cite{Mandel, Reid1989}. Two-mode (non-degenerate) squeezed states have two frequency modes which are entangled together, such that their noise is highly correlated. If the states are Gaussian, they can be succinctly described using a symplectic formalism \cite{Braunstein2005}, using the shared covariance matrix of the single-mode quadrature components of all different modes. The non-classicality in that case is encoded in the common fluctuations of some linear combination of the quadrature amplitudes, as shown for Gaussian states in the well-known criterion of Simon and Duan et al. \cite{Duan2000, Simon2000}, and generalized using higher moments by others \cite{Shchukin2005, Hillery2006, Agarwal2005}. 
	
	The Simon-Duan criterion for inseparability of states gives a simple constraint on the covariance (noise) of a Gaussian two-mode state \cite{Simon2000, Duan2000}. It gives a geometric interpretation of inseparability as a mirror transformation in phase-space, and gives a simple testable inequality for the covariance matrix, which is very useful in practice. The Simon-Duan criterion can be given a very different description and derivation using the quadrature powers. \textcolor{black}{Additionally, the moments used for the criterion are usually obtained using coherent homodyne detection \cite{Ou1992, Li2019}, but the direct relation to the quadrature powers reveals a completely different way to measure them,} using direct power measurements \cite{Shaked2018}, as illustrated in Fig. \ref{fig:phomodyne}a.
	
	The two-mode formalism unites the criteria for single-mode squeezing and two-mode inseparability, and puts them both on the same geometric interpretation. The bipartite non-separability is encoded in the attenuation of one complex quadrature power, and amplification of the the other, completely analogous to the degenerate case, and can be conveniently measured using direct power detection, as we discuss in Sec. \ref{sec:experimental_significance}.
	
	We define the covariance matrix encoding the correlations between the complex quadratures. This is different and lower in dimension than the full covariance matrix which explicitly considers the shared covariance of all single-mode quadrature components. Here, we only consider the Hermitian operators, nonetheless, it captures the full features of entangled two-mode $SU(1,1)$ states. 
	\begin{equation}
		V_{ij} = \left\langle \left\{ \Delta\hat{X}_{i}^{\dagger}, \Delta\hat{X}_{j}\right\} \right\rangle = \left\langle\hat{\Gamma}_{ij}\right\rangle - \text{Re}\left\{ \left\langle\hat{{X}}_i\right\rangle^*\left\langle\hat{X}_j\right\rangle \right\}.
	\end{equation}
	where $\Delta\hat{X}_{i}\equiv\hat{X}_{i}-\langle\hat{X}_{i}\rangle$. For concreteness, we discuss here only first and second-order correlations, which are sufficient for Gaussian states. More general inequalities can be obtained by considering higher moments of the complex quadratures $V_{ij}^{(nm)}=\left\langle \left\{ \Delta\hat{X}_{i}^{\dagger n},\Delta\hat{X}_{j}^{m}\right\} \right\rangle$.
	The elements of the covariance matrix capture the overall noise of the total complex amplitude, as illustrated in Fig. \ref{fig:quadrature_power_diagram}(a).
	For zero-mean states, e.g. two-mode squeezed vacuum state, the covariance matrix exactly equals the mean quadrature powers, as illustrated in Fig. \ref{fig:quadrature_power_diagram}(b) and \ref{fig:quadrature_power_diagram}(c). 
	
	The uncertainty principle gives a constraint on the covariances \cite{Simon1994, Weedbrook2012}. \textcolor{black}{First, let us write the covarince matrix in the basis of $\vec{X} = (\hat{X}_1, \hat{X}_2^\dagger)$. Noting that $V_{ij} = \left \langle \left[ \Delta\hat{X}_i^\dagger, \Delta{X_j} \right] \right \rangle + \langle \Delta\hat{X}_i^\dagger \Delta{X_j} \rangle $, which is nothing but the commutator written in matrix form.}
	\begin{equation}
		V +  
		\begin{pmatrix}
			0 & i \\
			-i & 0
		\end{pmatrix} \succeq 0
		\label{eq:semi-positiveness}
	\end{equation}
	\textcolor{black}{where the $\succeq$ sign implies that the sum of matrices on the left is a positive-semidefinite matrix,} which equivalently constraints the diagonal terms of the covariance matrix. This in turn gives the usual Heisenberg-like relations for the quadrature powers. This implies that $V$ is also positive semi-definite. \textcolor{black}{Since the matrix on the left is positive-semidefinite, its determinant is non-negative and hence its diagonal elements always satisfy:}
	\begin{equation}
		V_{11}V_{22} \ge 1
		\label{eq:var_h-rel}
	\end{equation}
	giving the usual Heisenberg-like relations for the quadrature powers.
	Graphically, this is reminiscent of the uncertainty area of the single-mode quadratures, only for the powers of the complex quadratures. In other words, the uncertainty principle for two-mode states can be defined in terms of the quadrature powers. \textcolor{black}{This is a restatement of the standard conditions on the covariance matrix, see e.g. the review by Braunstein et al. \cite{Braunstein2005} and by Weedbrook et al. \cite{Weedbrook2012}.}
	
	The Simon-Duan criterion \cite{Duan2000,Simon2000} states that a state is non-classical if the smallest eigenvalue of the covariance matrix is below unity. \textcolor{black}{Physically, when applying Eq.\ref{eq:semi-positiveness} to a general squeezed state, we find that the quadrature power decreases below the vacuum value (being 1 in our units) for some measurement phase $\phi$,}
	\begin{equation}
		V_{22}^{(\phi)} = V_{11}^{(\phi + \pi/2)} < 1.
	\end{equation}
	\textcolor{black}{Note that this is a specific example and nothing constrains $V_{22}$ to be quadrature power that is decreased below the vacuum level of 1. Generally, for different squeezing and measurement phases we could see different quadratures decrease below the vacuum level.}
	
	In the case of a zero-mean state, \textcolor{black}{this simplifies to the EPR criterion by simply plugging in the complex quadratures in terms of the single-mode quadratures.}
	\begin{equation}
		\left\langle \left(\Delta \left(\hat{y}_{+}+\hat{y}_{-}\right) \right)^2 \right\rangle + \left\langle \left(\Delta\left(\hat{x}_{+}-\hat{x}_{-}\right)\right)^2 \right\rangle < 1
	\end{equation}
	This has an important practical merit in that inseparability can be directly measured with power detection, rather than coherent homodyne. This has a number of benefits -- First, it allows effectively infinite measurement bandwidth and can simplify experimental schemes considerably, since no additional homodyne measurement is required. Additionally, many quantum optical platforms already use parametric amplifiers for readout and their measurement chains could be simplified by the use of homodyne measurements.

	\textcolor{black}{Since the complex quadratures have the same commutation relations as the single-mode quadratures, our covariance matrix has very similar structure, and many known results can be directly applied to it}.  For example, various important quantities of Gaussian states can be easily derived from the covariance matrix, such as the entanglement entropy of the state and other types of entropic measures \cite{Weedbrook2012, Braunstein2005}. One particularly convenient measure is the Gaussian R\'enyi entropy of order 2 \cite{Adesso2012} $\mathcal{S}_2 = \frac{1}{2}\ln\det V$, which is very simply related to the covariance matrix of a Gaussian state. Additionally, it can be directly related to the steerability of the state \cite{Kogias2015}.
	
	\subsection{Experimental significance and applications}
	\label{sec:experimental_significance}
	\textcolor{black}{Quantum sources of squeezed light can easily be very broadband, as exemplified by spontaneous down-conversion which can readily span up to $>100$ THz. While classical optical processing efficiently exploits the optical bandwidth resource, as represented for example in the frequency comb revolution of precision measurement \cite{Bloom2014, Wilken2012, Gohle2005}, quantum information processing is very inefficient in exploiting this natural bandwidth resource: despite the intrinsic availability of highly multimode optical fields that carry large quantities of quantum information, most applications to date are practically single-mode. The limit that prevents broadband multimode quantum information is the coherent detection of phase by homodyne measurement. To extract the optical phase, standard homodyne relies on balanced detection of the interference intensity, which is limited by the electronic bandwidth of the photo-detectors, and is therefore inherently single-mode \cite{Shaked2018}. In accordance, the standard theoretical treatment of squeezed light is solely relevant to narrowband degenerate squeezing, whereas the much more common case of two-mode squeezing is treated in a completely different and ungainly manner with narrowband correlations between the different frequencies. As a result, the usable bandwidth in current experiments is limited to a tiny fraction of the available squeezing. Multi-mode quantum detection in standard experiments requires multiple homodyne detectors \cite{Chen2014, Menicucci2008}, which is technically demanding, and all major work so far with multimode quantum light relied on detection of a single, or only a few modes. To harvest the orders-of-magnitude enhancement in processing throughput that the optical bandwidth enables, a paradigm shift is necessary in terms of both detection methods and theoretical analysis, which will open new avenues for quantum applications in sensing, communications and photonics-based quantum computation. Our formalism easily accommodates the measurement and characterization of broadband spectral quantum correlations and squeezing over an arbitrary optical bandwidth.}
	
	\begin{figure}
		\centering
		\includegraphics[width=8.6cm]{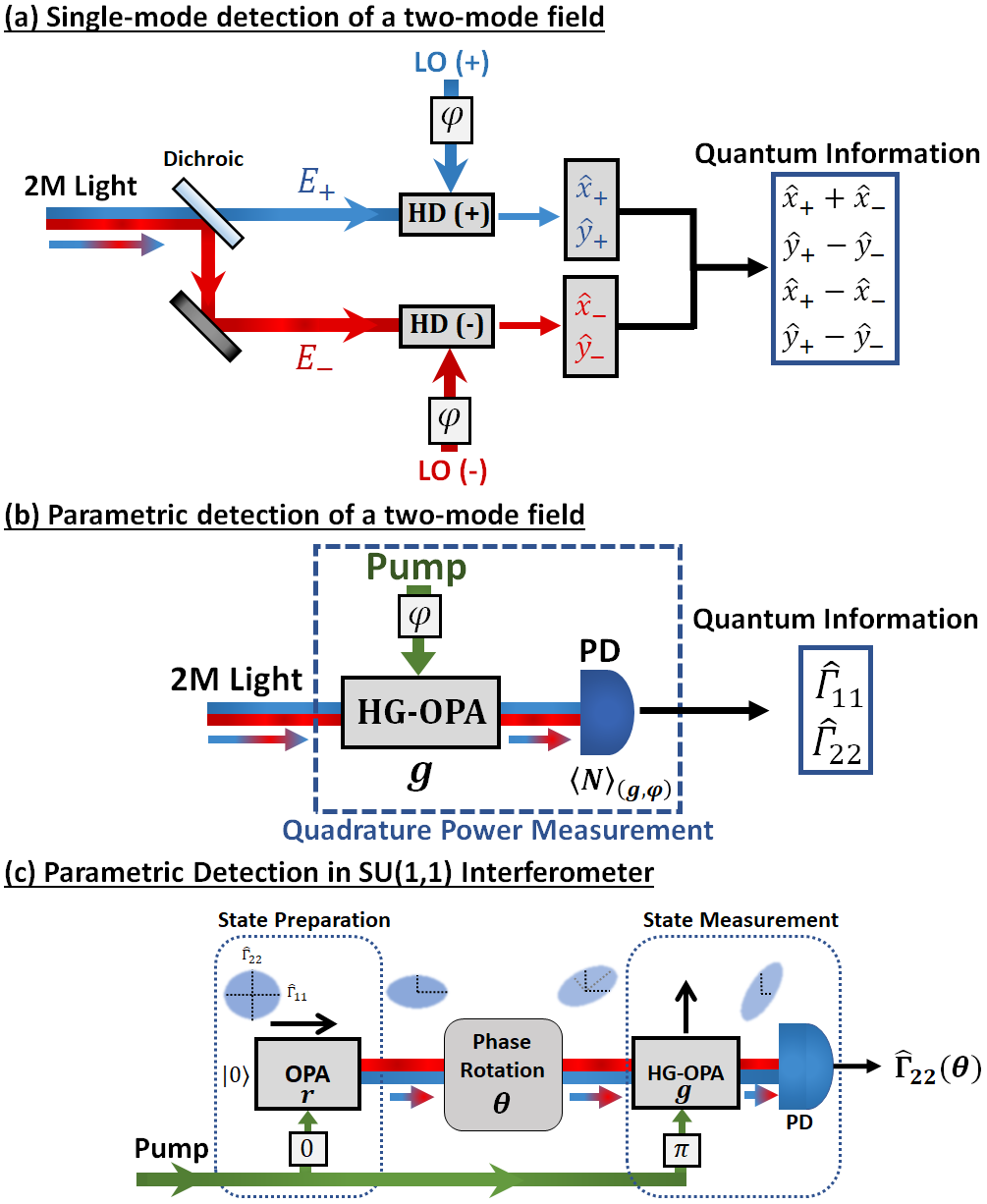}
		\caption{Detection alternatives of two-mode fields. LO: Local oscillator. \textbf{HD}: Homodyne detection. \textbf{HG-OPA}: High-gain optical parametric amplifier. \textbf{PD}: Photodetection. \textbf{(a)} Standard detection of quantum squeezing based on balanced homodyne detection. \textbf{(b)} Detection of the quadrature power based on high-gain parametric amplification and direct detection. \textbf{(c)} The $SU(1,1)$ interferometer can be seen as preparation of a phase-dependent state with squeezing ratio $r$, and then measurement of one quadrature power using another high-gain parametric amplifier with gain $g$. The blue ellipses indicate the noise of the two-mode state at each stage of the interferometer.}
		\label{fig:phomodyne}
	\end{figure}
	
	We employ the formalism to describe how the quadrature powers make a very useful observable for broadband quantum states. We expand on previous work, such as measurement of the complex quadratures \cite{Li2019} and the quadrature powers \cite{Shaked2018}, as well as $SU(1,1)$ interferometry \cite{Chekhova2016}. The results presented here are not new, but they were never described in this theoretical framework and are illuminated in a new light by the complex quadratures formalism. We define the expectation value of a state after being parametrically amplified, normalized by the expectation value of a squeezed vacuum state:
	\begin{equation}
		\label{eq:squeezed_value}
		\left\langle\hat{O}\right\rangle_{(g)} = \frac{\left\langle \hat{S}^\dagger(g)\hat{O} \hat{S}(g) \right\rangle}{\left\langle 0 \rvert \hat{S}^\dagger(g)\hat{O} \hat{S}(g) \lvert 0 \right\rangle}
	\end{equation}
	where $g = |g|e^{i\phi}$ refers to a parametric amplifier with parametric gain $g$ and phase $\phi$. There are often two types of parametric gain in our schemes -- $r$ which is the gain of the squeezer, or the squeezing ratio of the input state, and $g$ which denotes the parametric gain of the parametric amplifier used for state measurement, and can be assumed much larger than $r$ for the sake of the current discussion of a robust and precise homodyne measurement. Using Eq. \ref{eq:squeezed_value}, we find a direct relation between the output power after a parametric amplifier and the quadrature powers at the input:
	\begin{equation} 
		\label{opout}
		\left<\hat{N}\right>_{(g)} \equiv \frac{1}{\sinh 2g} \left[ \left\langle \normord{\hat{\Gamma}^{(\phi)}_{11}} \right\rangle e^{2g} + \left\langle \normord{\hat{\Gamma}^{(\phi)}_{22}} \right\rangle e^{-2g}\right].
	\end{equation}
	\textcolor{black}{where again, $\phi$ is the squeezing phase, determined by the pump.}
	This indicates that for sufficient parametric gain $g$, the amplified quadrature power dominates the overall power, allowing direct measurement of the quadrature powers by tuning the phase of the pump. \textcolor{black}{A derivation of Eq. \ref{opout} is given in Appendix \ref{sec:derivation_squeezed_values}.}
	
	If the measurement gain $g$ is not large enough, the output power would include contributions from both quadrature powers. Importantly, this only requires the parametric gain to be sufficiently large, but not that the amplifier behaves as an ideal squeezer, which greatly relaxes the experimental requirements on the parametric amplifier.
	
	Measurement of the quadrature powers have a number of benefits. First, for various applications, full measurement of the complex quadratures is not required and it is often enough to measure only the quadrature powers, since they generate and completely describe states with $SU(1,1)$ symmetry. A second key point is that since the quadratures are slowly-varying, the measurement bandwidth is increased by several orders of magnitude, limited only by the phase-matching bandwidth of the HG-OPA, as was shown in previous work \cite{Shaked2018}. 
	\textcolor{black}{This is an important point that we would like to emphasize - since the quadrature amplitudes are independent of bandwidth, they can be measured directly using power measurements and spectral analysis. That point is discussed in more depth in the Appendix \ref{sec:multimode_quadratures}}.
	
	An experimental scheme that is closely related to this terminology is the $SU(1,1)$ interferometer \cite{Lett_2017, Kalashnikov2016}. The two-mode formalism can greatly simplify the description of such devices \cite{michael_2019}. In $SU(1,1)$ interferometers, the beam-splitters are replaced by two parametric amplifiers in series, pumped by the same pump with a small phase-shift induced between them. The amplification axes and gain of the two amplifiers are arranged to completely null each other, such that light will be detected at the output of the second amplifier only in the presence of a non-zero phase shift. The second amplifier then acts to measure the quadrature orthogonal to the one initially amplified. Using a high-gain parametric amplifier, this coincides with measurement of:
	\begin{equation}
		\hat{\Gamma}_{22} \rightarrow \sin^{2}\theta e^{2r} \hat{\Gamma}_{11} + \cos^{2}\theta e^{-2r} \hat{\Gamma}_{22} \\
		- \sin\theta\cos\theta\hat{\Gamma}_{12}. 
	\end{equation}
	\textcolor{black}{As illustrated in Fig. \ref{fig:phomodyne}, $r$ is the gain of the first amplifier, while $\theta$ is the phase difference accrued between the two amplifiers.} Note that $\hat{\Gamma}_{12}$ enters as interference term, justifying its definition as a coherence function. If the input state is a vacuum state, the output quadrature power simplifies to
	\begin{equation}
	    \label{eq:su11_out}
		\left\langle \hat{\Gamma}_{22} \right\rangle = \frac{1}{2} \left[ \cosh {2r} - \cos(2\theta)\sinh {2r} \right].
	\end{equation}
	If no external phase $\theta$ is present, the output power falls below the input level, and exhibits a minimum value. If some phase shift is present, then the cancellation will no longer be complete and the output will rise above the minimum level. In the perspective of the complex quadrature, the $SU(1,1)$ interferometer is seen as simply the manipulation and measurement of the quadrature powers, rather than an interferometer. We note that a similar perspective was offered by Caves et al. in a recent work \cite{Caves2020}.
	
	Fundamental quantities of the $SU(1,1)$ interferometer, like power and visibility, are directly related to the basic quantities in the complex quadrature formalism. The visibility of the interferometer defined as the difference between the maximum and minimum powers, can also be calculated based on our formalism. In the limit of high-gain, where the minimum and maximum values correspond to the squeezed and stretched quadratures, respectively:
	\begin{subequations}
		\begin{align}
			\mathcal{W}_1 \equiv \frac{\left\langle \hat{\Gamma}^{(0)}_{11} \right\rangle-\left\langle \hat{\Gamma}^{(0)}_{22} \right\rangle}{\left\langle\hat{\Gamma}_{11}\right\rangle + \left\langle\hat{\Gamma}_{22}\right\rangle} = \frac{\left\langle \hat{K}_{1} \right\rangle}{\left\langle \hat{K}_3 \right\rangle} \\
			\mathcal{W}_2 \equiv \frac{\left\langle \hat{\Gamma}^{(\pi/4)}_{11} \right\rangle-\left\langle \hat{\Gamma}^{(\pi/4)}_{22} \right\rangle}{\left\langle\hat{\Gamma}_{11}\right\rangle + \left\langle\hat{\Gamma}_{22}\right\rangle} = \frac{\left\langle \hat{K}_{2} \right\rangle}{\left\langle \hat{K}_3 \right\rangle}
		\end{align}
	\end{subequations}
	We see that the two-photon visibilities are directly proportional to the generators of squeezing, and together with the overall power they determine the state. It is also interesting to note that the single particle visibility is given by the generators of $SU(2)$ \cite{Yurke1986}, \textcolor{black}{given here for completeness},
	\begin{subequations}
	    \begin{align}
	        \hat{J}_1 &= \frac{1}{2} \left ( \hat{a}_+^\dagger \hat{a}_- + \hat{a}_-^\dagger \hat{a}_+\right) \\
	        \hat{J}_2 &= -\frac{i}{2} \left ( \hat{a}_+^\dagger \hat{a}_- - \hat{a}_-^\dagger \hat{a}_+\right) \\
	        \hat{J}_3 &= \frac{1}{2} \left ( \hat{a}_+^\dagger \hat{a}_+ - \hat{a}_-^\dagger \hat{a}_-\right) \\
	        [\hat{J}_i, \hat{J}_j] &= i\epsilon_{ijk} \hat{J}_k \\
	        J^2 &= \frac{N}{2} \left [ \frac{N}{2} + 1 \right]
	    \end{align} 
	\end{subequations}
	\textcolor{black}{For example, the single-particle visibility in an interferometric experiment could be given by}
	\begin{equation}
	    \mathcal{V}_1 = \frac{\left\langle \hat{J}_1 \right\rangle}{\left\langle \hat{K}_3 \right\rangle}.
	\end{equation}
	In general, the single-particle and two-particle visibilities do not commute, and satisfy a complementarity relation \cite{franson_1989, Jaeger1993, Peled2020}, 
	\begin{equation}
		\mathcal{V}_1^2 + \mathcal{W}_1^2 \le 1.
	\end{equation}
	Thus, a state that shows high visibility in a single-particle ($SU(2)$) interference experiment, would show low visibility in a two-particle $SU(1,1)$ interference experiment, and vice versa.
	
	\section{Discussion}
	\label{sec:conclusions}
	We presented a two-mode formalism for two-mode devices, where the Hermitian quadrature amplitudes of the single-mode field are generalized to a set of complex non-local operators. Their components are directly related to the EPR observables for squeezed-states, and they cannot be ascribed to any of the comprising single modes, but only to the overall two-mode field. In that formalism, the complex quadrature operators change very simply under the multimode Bogoliubov transformations characterizing parametric devices, and degenerate and non-degenerate parametric phenomena are described on equal footing within the same framework.
	In addition, we defined a set of Hermitian operators quadratic in the two-mode quadrature operators, which we termed the quadrature power and coherence operators. We have shown how these operators can be used to define all relevant observables of a two-mode field, and that their algebra naturally generates the $SU(1,1)$ algebra of two-mode devices. Importantly, our formalism provides a very useful description for broadband light, and suggests generalized observables that characterize the entire broadband field.
	
	Furthermore, we have shown how they directly correspond to the observable quantities of an $SU(1,1)$ interferometer, like visibility and power. We demonstrated how coherent properties of two-mode fields (which are usually assumed to require coherent homodyne detection) can be reduced to direct measurements of the quadrature powers. This also leads us to advocate a different point of view on $SU(1,1)$ interferometers, as measurement of the displaced complex quadratures. The important property of this type of measurement, which is detected through the power and coherence quadrature operators, is that it encapsulates the phase-dependent correlations between the modes in a very efficient manner. These correlations carry the quantum signature of squeezed states, and being able to measure them efficiently is important for many applications, such as sensitive displacement detection \cite{michael_2019} and quantum illumination receivers \cite{Zhang2015, Zhuang2017}.

\section{Acknowledgements}
    We thank our lab members and David Karasik for fruitful discussions.
    \emph{Funding} E.C. and A.P. were supported by the Israel Innovation authority under grant 70002. E.C. acknowledges support from the Quantum Science and Technology Program of the Israeli Council of Higher Education and from FQXi and from the Pazy foundation.

	\bigskip
	\noindent
	
	\bibliography{bibliography}
	
	\section*{Appendix}
	\label{sec:appendix}
	
	\subsection{Derivation of the squeezed expectation values}
	\label{sec:derivation_squeezed_values}
	\textcolor{black}{Eq. \ref{opout} is derived directly from Eq.} \ref{eq:squeezed_value} 
	\begin{equation*}
		\left\langle\hat{O}\right\rangle_{(g)} = \frac{\left\langle \hat{S}^\dagger(g)\hat{O} \hat{S}(g) \right\rangle}{\left\langle 0 \rvert \hat{S}^\dagger(g)\hat{O} \hat{S}(g) \lvert 0 \right\rangle}	    
	\end{equation*}
	\textcolor{black}{by using the relation in Eq. \ref{eq:total_power}, written here again for convenience.}
	\begin{equation*}
	    \hat{N} = \normord{\hat{\Gamma}_{11}} + \normord{\hat{\Gamma}_{22}}.
	\end{equation*}
	\textcolor{black}{The normal ordering is simply an aesthetic way to remove the constant vacuum value, and does not change how the quadrature power operators transform. The transformation of the quadrature powers under squeezing is given by Eq. \ref{eq:quadrature_powers_squeezing}, and the operators simply scale the by the appropriate squeezing factor,}
	\begin{equation*}
	      \normord{\hat{\Gamma}_{11}} + \normord{\hat{\Gamma}_{22}} \rightarrow \normord{\hat{\Gamma}_{11}}e^{2r} + \normord{\hat{\Gamma}_{22}}e^{-2r}
	\end{equation*}
	 \textcolor{black}{By the definition of the squeezed expectation value, we normalize by calculating the mean number of photons in a squeezed vacuum state,}
	 \begin{equation*}
	     {\langle 0 | \hat{N} |0 \rangle}_{(g)} = \sinh g,
	 \end{equation*}
	\textcolor{black}{which gives us exactly Eq. \ref{opout}.}
	
	\subsection{Derivation of the SU(1,1) interferometer}
	\textcolor{black}{As illustrated in Fig. \ref{fig:phomodyne}, the $SU(1,1)$ interferometer is comprised of three steps. First, we generate a squeezed vacuum state. In the Heisenberg picture, this corresponds to a simple scaling transformation of the quadrature amplitude $\hat{\Gamma}_{11} \rightarrow \hat{\Gamma}_{11}e^{2r}$. This is followed by a rotation transformation, due to the phase between the amplifiers. In the scheme, we measure the attenuated quadrature, which is transformed according to}
	\begin{multline*}
	    \hat{\Gamma}_{22}\rightarrow\left(-\hat{X}_{1}\sin\theta+\hat{X}_{2}\cos\theta\right)\left(-\hat{X}_{1}^{\dagger}\sin\theta+\hat{X}_{2}^{\dagger}\cos\theta\right) \\
	    \rightarrow\sin^{2}\theta e^{2r}\hat{\Gamma}_{11}+\cos^{2}\theta e^{-2r}\hat{\Gamma}_{22}-\sin\theta\cos\theta\hat{\Gamma}_{12}.
	\end{multline*}*
    
    \textcolor{black}{Taking the expectation value on the input vacuum state finally gives Eq. \ref{eq:su11_out}, by using the following vacuum expectation values,}
	\begin{subequations}
	    \begin{align}
	        \left\langle \hat{\Gamma}_{22}^{2}\right\rangle &= \left\langle \hat{\Gamma}_{11}^{2}\right\rangle = 1/2 \\
	        \left\langle \hat{\Gamma}_{12}^{2}\right\rangle &= 0
	    \end{align}
	\end{subequations}

	\subsection{Multimode fields and continuum description}
	\label{sec:multimode_quadratures}
	One of our goals in the definition of the complex quadratures is to give a description of broadband multimode fields, and identify their relevant global observables. Our definitions above can be easily extended to a continuum description, where different modes are now denoted by an argument rather than a subscript. The carrier dependence is suppressed, and all frequencies are assumed to be around the carrier frequency $\Omega$. The mode operators now satisfy the continuum commutation relations
	\begin{subequations}
		\label{eq:continuum_commutators}
		\begin{align}
			\left[\hat{a}^\dagger(\epsilon), \hat{a}(\epsilon')\right] &= 2\pi \delta(\epsilon - \epsilon') \\
			\left[\hat{a}(\epsilon), \hat{a}(\epsilon')\right] &= 0
		\end{align}
	\end{subequations}
	And similarly for the quadrature operators
	\begin{subequations}
		\label{eq:complex_quadratures_continuum_1}
		\begin{align}
			\hat{X}_{1}\left(\epsilon\right)=\frac{1}{\sqrt{2}}\left(\hat{a}(\epsilon)+\hat{a}^{\dagger}(- \epsilon)\right) \\
			\hat{X}_{2}\left(\epsilon\right) = \frac{1}{\sqrt{2}i}\left(\hat{a}(\epsilon)-\hat{a}^{\dagger}(- \epsilon)\right) 
		\end{align}
	\end{subequations}
	\begin{subequations}
		\label{eq:complex_quadratures_continuum_commutators}
		\begin{align}
			\left[\hat{X}_1(\epsilon), X_2^\dagger(\epsilon')\right] &= \left[\hat{X}_1^\dagger(\epsilon), X_2(\epsilon')\right] = 2\pi i \delta(\epsilon - \epsilon')\\
			\left[\hat{X}_1(\epsilon), X_1^\dagger(\epsilon')\right] &= \left[\hat{X}_2(\epsilon), X_2^\dagger(\epsilon')\right] = 0 \\
			\left[\hat{X}_1(\epsilon), X_2(\epsilon')\right] &= 0
		\end{align}
	\end{subequations}
	The quadrature power operators now become symmetric functions of $\epsilon$, representing the quadrature power spectral density
	\begin{subequations}
		\begin{align}
			\left[\hat{\Gamma}_{11}(\epsilon), \hat{\Gamma}_{22}(\epsilon')\right] &= 4\pi i \hat{\Gamma}_{12}(\epsilon) \delta(\epsilon - \epsilon') \\
			\left[\hat{\Gamma}_{22}(\epsilon), \hat{\Gamma}_{12}(\epsilon')\right] &= -4 \pi i \hat{\Gamma}_{22}(\epsilon) \delta(\epsilon - \epsilon')  \\
			\left[\hat{\Gamma}_{12}(\epsilon), \hat{\Gamma}_{11}(\epsilon')\right] &= -4 \pi i \hat{\Gamma}_{11}(\epsilon) \delta(\epsilon - \epsilon').
		\end{align}
	\end{subequations}
	The entire global quadrature power can be measured in a simple, single-measurement scheme of a strong parametric amplification followed by direct photo-detection. We define the total quadrature power to be
	\begin{equation}
		\hat{{\varGamma}}_{ii} = \int_{-\infty}^{\infty} \frac{d\epsilon}{2\pi} \hat{\Gamma}_{ii}(\epsilon) 
	\end{equation}
	the total power is of course given still by tracing over the total quadrature power, \textcolor{black}{and the quadrature power related to each pair can be obtained by measuring the power spectral density of the field.}

\end{document}